\documentclass[aps,preprint,showpacs]{revtex4}
\usepackage{graphicx}
\usepackage{amssymb}

\usepackage{longtable}
\usepackage{float}
\usepackage{amsmath}

\begin{document}

\title{Kinetic roughening-like transition with finite nucleation barrier}

\author{James F. Lutsko}

\affiliation{Center for Nonlinear Phenomena and Complex Systems CP 231, Universit\'{e}
Libre de Bruxelles, Blvd. du Triomphe, 1050 Brussels, Belgium}

\email{jlutsko@ulb.ac.be}

\homepage{http://www.lutsko.com}

\author{Vasileios Basios}

\affiliation{Center for Nonlinear Phenomena and Complex Systems CP 231, Universit\'{e}
Libre de Bruxelles, Blvd. du Triomphe, 1050 Brussels, Belgium}

\author{Gr\'{e}goire Nicolis}

\affiliation{Center for Nonlinear Phenomena and Complex Systems CP 231, Universit\'{e}
Libre de Bruxelles, Blvd. du Triomphe, 1050 Brussels, Belgium}

\author{John J. Kozak}
\affiliation{Dept. of Chemistry, DePaul University, 243 South Wabash Avenue, Chicago, Illinois 60604, USA}

\author{Mike Sleutel and Dominique Maes}

\affiliation{Flanders Interuniversity Institute for Biotechnology (VIB),Vrije Universiteit Brussel, 
Pleinlaan 2 Building E, 1050 Brussel, Belgium}

\pacs{82.20.Wt, 81.10.Aj, 82.60.Nh}

\begin{abstract}
Recent observations of the growth of protein crystals have identified
two different growth regimes. At low supersaturation, the surface of
the crystal is smooth and increasing in size due to the nucleation
of steps at defects and the subsequent growth of the steps. At high
supersaturation, nucleation occurs at many places simultaneously,
the crystal surface becomes rough and the growth velocity increases
more rapidly with increasing supersaturation than in the smooth regime.
Kinetic roughening transitions are typically assumed to be due
to the vanishing of the barrier for two-dimension nucleation on the
surface of the crystal. We show here, by means of both analytic mean
field models and kinetic Monte Carlo simulations that a transition
between different growth modes reminiscent of kinetic roughening can also arise as a kinetic effect occurring at finite
nucleation barriers. 
\end{abstract}

\date{\today }

\maketitle

\section{Introduction}

Crystal growth takes place at the interface of the solid and liquid
phases and is determined by the structure of the crystal surface, the
temperature and the supersaturation. Each growth unit is incorporated
into the crystal when the energy barrier for absorption is overcome by
the unit's kinetic energy. For rough surfaces, such as K- and S-faces,
the density of privileged sites, i.e. kinks, is high and growth is fast.
In this work, however, we consider initially flat faces such as the F-faces. For
these surfaces, at low supersaturation values, kink sites are scarce and
the growth process is slow. Growth units encounter the kinks directly
from solution, or via a random walk due to 2D-diffusion on the surface.
New steps are either generated through spiral dislocations or
2D-nucleation\cite{Saito,mike-doc-04,mike-doc-05,mike-doc-06,mike-doc-07,mike-doc-08}. The temperature is a determining factor. It is
well known that there exists a critical value, $T_{R}$, where
the step edge free energy vanishes and the crystal surface becomes
rough at the growth unit scale. This is a thermodynamic phase transition,
known as \emph{thermal} roughening\cite{Cabrerra,mike-doc-10,mike-doc-11,mike-doc-12,mike-doc-13,mike-doc-14}. 

At constant temperature, the driving force is the chemical potential
difference or supersaturation and an increase
in the driving force will translate into a decrease in critical nucleus
size. Since for temperatures less than $T_{R}$ the step edge's free
energy is greater than zero, according to the Gibbs-Thomson relation
\cite{mike-doc-15,mike-doc-16}, an activation barrier for the formation
of the critical two-dimensional nucleus exists and with it, a critical
nucleus size larger than the crystal\textquoteright{}s individual
building blocks. The size of this 2D-critical nucleus is inversely
proportional to the temperature and the supersaturation \cite{Saito}.

Recently, it has been  reported for protein crystallization systems
\cite{mike-doc-30,Mike1} that for temperatures lower than $T_R$ there exists a critical supersaturation for which the size of
the critical two-dimensional nucleus is reduced to the order of one
growth unit and the activation barrier for two-dimensional nucleation
essentially vanishes. Due to a large step density and a very small
two-dimensional critical nucleus at elevated supersaturations, the
surface becomes rough and offers many favorable sites almost uniformly
distributed across the surface. Consequently, arriving molecules can
be incorporated quasi at any site. This transition from a slow, layer-by-layer
growth regime to a fast continuous growth regime at high driving forces
is referred to as \emph{kinetic} roughening transition \cite{mike-doc-17,mike-doc-18,mike-doc-19,mike-doc-20,mike-doc-21,mike-doc-22}. 

Kinetic roughening for the case of crystallization from solution has
been observed for many small molecules, i.e. $SiO_{2}$, $Al_{2}O_{3}$,
$ZnO$ and $ZnS$ \cite{mike-doc-25,mike-doc-26}, n-paraffins \cite{mike-doc-27,mike-doc-28}
and many others. For the case of macromolecules such as in protein
crystallization, kinetic data of continuous growth has only
been presented for lysozyme \cite{mike-doc-29,mike-doc-30,mike-doc-31}.
Very recently, quantitative
data for both the classical layer-by-layer growth mechanism and the
kinetic roughening regime for the crystallization of glucose isomerase
from \textit{Streptomyces rubiginosus}  have been presented \cite{Mike1}. 

In this paper we show that kinetic roughening-like transitions
in the presence of a \emph{finite} nucleation barrier can occur owing to kinetic mechanisms still
rendering possible the transition from slow layer-by-layer growth regime
(low supersaturations) to a fast continuous growth regime (high supersaturations). The general scheme used views the process of crystal
growth at constant temperature driven by the difference in chemical
potential of the solution around the crystal surface and 
the reservoir of the surrounding liquid phase. The local
environment around the surface incorporates the barriers, kinks and
dislocations while relevant kinetic processes such as absorption,
surface diffusion and nucleation contribute to the growth. A mean field model and a Monte Carlo simulation are proposed and serve
as an illustrative generic explanation of the observed data from protein
crystallization \cite{Mike1}.
\begin{table}[H]
\begin{longtable}{|l|c|c|c|c|}
\cline{1-1} \cline{3-3} \cline{5-5} 
Large reservoir  & $\leftrightharpoons$ & \multicolumn{1}{c|}{Local environment } & $\underset{desorption}{\rightleftharpoons}$ & Surface: absorption, surface diffusion, \tabularnewline
\newpage &  & around surface &  & nucleation, growth\tabularnewline
\cline{1-1} \cline{3-3} \cline{5-5} 
\newpage
\end{longtable}
\end{table}

In Section II a minimal
mean field model is developed incorporating cluster formation and possessing the
two main asymptotic regimes, exhibiting the transition from
low to high supersaturation growth. In Section III we investigate
an extension of a {}``solid to solid'' (SOS) Monte Carlo simulation
linked to, but more general than, the mean field model, and accounting
for kinetic roughening in the presence of finite nucleation barrier
and dislocations. The main conclusions are summarized in Section IV.

\section{Mean Field Model}

Our mean-field model of crystal growth is based on a generalization
of the Burton-Cabrera-Frank model for epitaxial growth\cite{Cabrerra,Cabrera2,Cabrera3}
and is also motivated by theoretical work hinting at a competition
between surface aggregation and the formation of cluster by direct
incorporation of material from solution\cite{Nicolis}. Let $X_{0}$
and $X$ denote, respectively, molecules in the
bulk solution far from the growing crystal and in solution near the
crystal surface; and $Y$, $C_{N}$ the concentration of  molecules
adsorbed onto the surface but not incorporated into the solid and
of clusters constituting islands of solid material of size
$N$. There are four basic physical processes that we consider. First
is the diffusion of molecules between the bulk solution and the solution
near the growing crystal, $X_{0}\rightleftarrows X$. Distinguishing
between the bulk solution and that near the surface allows us to model
the formation of a depletion layer as molecules are adsorped onto
the surface and incorporated into the solid. The second process is
that of adsorption from the solution onto the surface, $X\rightleftarrows Y$.
We consider two pathways for the formation of an island of size $N$:
adsorption and aggregation, $X+\left(N-1\right)Y\rightleftarrows C_{N}$,
and aggregation alone, $NY\rightleftarrows C_{N}$. Denoting the concentrations
of $X$ and $X_{0}$ by $x$ and $x_{0}$ respectively, the surface
concentration of adsorbates and $N$-islands by $y$ and $c_{n}$,
and assigning rate constants to each process gives the rate equations
\begin{eqnarray}
\frac{dx}{dt} & = &
ax_{0}+k_{0}^{\prime}y-\left(a^{\prime}+k_{0}\right)x+k_{1}^{\prime}c_{N}-k_{1}xy^{N-1}
\label{model} \\
\frac{dy}{dt} & = & k_{0}x-k_{0}^{\prime}y-\left(N-1\right)k_{1}xy^{N-1}-Nk_{2}y^{N}+\left(\left(N-1\right)k_{1}^{\prime}+Nk_{2}^{\prime}\right)c_{N}\nonumber \\
\frac{dc_{N}}{dt} & = & k_{1}xy^{N-1}+k_{2}y^{N}-k_{1}^{\prime}c_{N}-k_{2}^{\prime}c_{N}\nonumber \end{eqnarray}

The rate constants $a$ and $a^{\prime}$ control the rate of exchange
between the bulk and the solution near the crystal surface (in effect,
this represents a zero-dimensional model for spatial diffusion). The
constants $k_{0}$ and $k_{0}^{\prime}$ control the rates of adsorption
and desorption. The remaining rate constants pertain to the rates
of nucleation and dissolution of clusters. Note that the probability to bring together $N$ molecules is taken to be proportional to the N-th power of the concentration: this is equivalent to assuming the rate goes like $e^{-\beta N \Delta W}$ where $\Delta W$ is the work of formation per molecule of a cluster of size $N$  and using the low-density approximation that the free energy of a molecule in solution is proportional to the log of the concentration.

Notice that in the perspective of Eqs.(\ref{model}) growth arises from
the fact that the number of clusters $c_{N}$ ($N$ given) increases
until the whole surface is covered. Thus, the total solid mass is $M=Nc_{N}$ and the growth velocity is $dM/dt = dNc_{N}/dt$. (In this zero-dimensional model, there is no distinction between growth within a layer and growth perpindicular to the surface.) In actual fact, in addition to
this mechanism one expects that the formation of bigger clusters from
smaller ones through, e.g., the process $C_{N}+Y \rightarrow C_{N+1}$
should also contribute to growth. In this Section, we will limit
ourselves to the model of Eqs.(\ref{model}), which will be used as a
reference for sorting out the different growth regimes in a
transparent manner and as a basis for comparison with the Monte Carlo
simulations. This approximation is expected to be valid for high supersaturations where step growth is less important. Finally, we only consider the process in which one molecule is directly incorporated from the solution since, for the conditions considered here, we expect this to be dominant over processes involving two or more molecules coming from solution. 

\subsection{A simplified model}

To illustrate the basic mechanism leading to a change in growth regimes,
we first ignore depletion of the protein in solution near the interface
so that $a=a^{\prime}=0$ and $x=x_{0}$ and neglect all dissolution
processes so that $k_{0}^{\prime}=k_{1}^{\prime}=k_{2}^{\prime}=0$. This limit should describe to a good approximation crystals growing in the kinetically-limited regime. What remains is then  the simplified dynamics,\begin{eqnarray}
\frac{dy}{dt} & = & k_{0}x-\left(N-1\right)k_{1}xy^{N-1}-Nk_{2}y^{N}\\
\frac{dc_{N}}{dt} & = & k_{1}xy^{N-1}+k_{2}y^{N}\nonumber \end{eqnarray}
Since there is no detailed balance, there is no equilibrium state.
However, a steady state where $\frac{dy}{dt}=0$ exists, implying a constant growth velocity,  $N\frac{dc_{N}}{dt}=v$. The concentration of adsorbed molecules
is then determined by
\begin{equation}
0=k_{0}x-\left(N-1\right)k_{1}xy^{N-1}-Nk_{2}y^{N}\end{equation}
and the velocity is \begin{equation}
v=Nk_{1}xy^{N-1}+Nk_{2}y^{N}\end{equation}
When the concentration of protein in solution is small compared to
that adsorbed on the surface, $x<<y$, one finds \begin{eqnarray}
y & = & \left(\frac{k_{0}x}{Nk_{2}}\right)^{1/N}+..., \label{result1} \\
v & = & k_{0}x+...\nonumber \end{eqnarray}
while the reverse circumstance, $x>>y$, gives \begin{eqnarray}
y & = & \left(\frac{k_{0}}{\left(N-1\right)k_{1}}\right)^{1/\left(N-1\right)}+... \label{result2} \\
v & = & \frac{N}{N-1}k_{0}x+...\nonumber \end{eqnarray}
There is therefore a crossover from small $x$ behavior, where the aggregation
process, $Ny\rightleftarrows c_{N}$, dominates with velocity $v=k_{0}x$
to the large $x$ behavior where the $x+\left(N-1\right)y\rightleftarrows c_{N}$
process dominates with velocity $v=\frac{N}{N-1}k_{0}x$. Since this transition occurs in the normal growth regime - below the usual roughening transition - the growth rates are, in these limits,  linear functions of the concentration, as expected. Small and
medium sized clusters are those contributing the most in the
switching, since for large $N$ the growth velocities of the two
regimes become practically indistinguishable. As we will discuss elsewhere, to get a more detailed sense of the interplay between these mechanisms, a simple
model can be  formulated for which numerically-exact results can be obtained using the theory of finite Markov processes\cite{Kozak}.

\subsection{Full model}
\begin{figure}
\begin{centering}
\resizebox{10cm}{!}{ {\includegraphics[height=10cm,angle=-90]{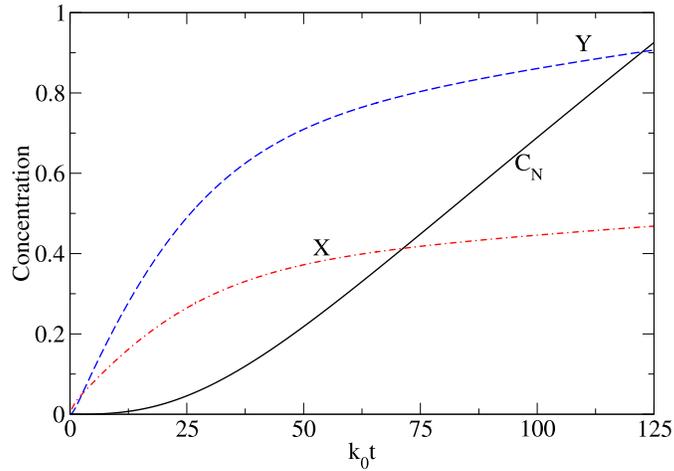}}} 
\par\end{centering}
\caption{Time-dependence of the quantities $x(t)$, $y(t)$ and
$c_{N}(t)$ as results from a numerical solution
of the full model, Eqs.(\ref{model}), with the parameters given in the text.}
\label{fig1} 
\end{figure}

\begin{figure}

\begin{centering}
\resizebox{10cm}{!}{ {\includegraphics[height=10cm,angle=-90]{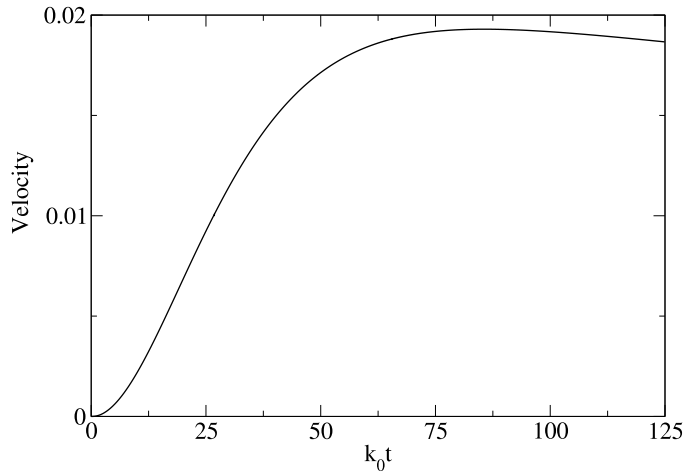}}} 
\par\end{centering}

\caption{The dimensionless growth velocity, $\frac{dNc_N}{dk_{0}t}$,  as a function
of time for the same parameters as used in Fig. \ref{fig1}. The supersaturation in this case is $log(x/x^{*}) \approx -0.84$.}

\label{fig2} 
\end{figure}

\begin{figure}
\begin{centering}
\resizebox{10cm}{!}{ {\includegraphics[height=10cm,angle=-90]{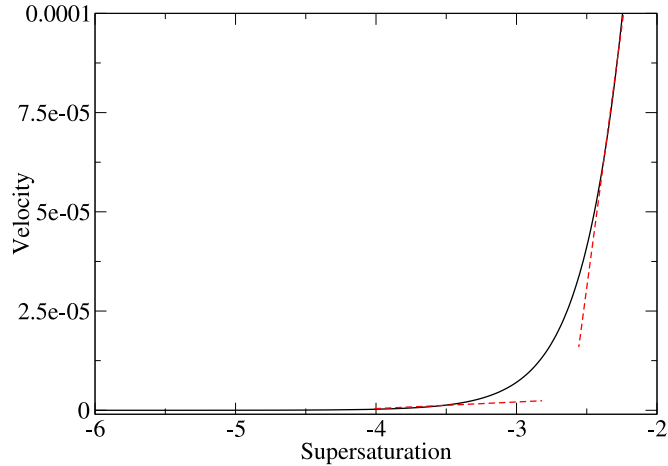}}} 
\par\end{centering}

\caption{Maximal growth velocity as a function of supersaturation, $log(x/x^{*})$, for the parameters
given in the text. The broken lines extrapolate the values into
asymptotic regions.}

\label{fig3} 
\end{figure}

In the simplified model, all dissolution processes were neglected. Experimental investigations\cite{Chen2002, Sleutel2008} have shown, however, that this is not valid for protein crystals growing at low driving forces. Here, we therefore incorporate desorption and dissolution processes in the full model. 
Including the desorption of surface molecules, $k_{0}^{\prime}\ne 0$, does not change the
picture dramatically: the small x behavior is no longer linear but
there is still a crossover between one well-defined growth mode at
small $x$ to another, faster mode at large $x$ as in the simple
model. Including all of the evaporation processes, $k_{1,2}^{\prime} \ne 0$, however, does change
the picture significantly. First, detailed balance defines an equilibrium
state at some finite value of $x_{0}$. For bulk concentrations above
this value, growth occurs but it will eventually end when a steady
(equilibrium) state is reached. In between, a quasi-steady growth
regime occurs during which a growth velocity can be defined. Detailed
numerical solutions of the model again show a crossover in the growth
velocity as the bulk concentration increases. The requirement of detailed
balance at equilibrium, i.e. that the forward rates and backward rates
of all processes are equal, gives \begin{eqnarray}
ax_{0} & = & a^{\prime}x^{*}\\
k_{0}x^{*} & = & k_{0}^{\prime}y^{*}\nonumber \\
k_{1}^{\prime}c_{N}^{*} & = & k_{1}x^{*}y^{*N-1}\nonumber \\
k_{2}y^{*N} & = & k_{2}^{\prime}c_{N}^{*}\nonumber \end{eqnarray}
 where $x^{*}$ is the equilibrium value of $x$, etc. Given a value
for the concentration in the bulk solution, these relations determine
the equilibrium values of all of the other quantities as well as imposing
the consistency relation 
\begin{equation} \label{balance}
k_{0}^{\prime}k_{1}k_{2}^{\prime}=k_{0}k_{1}^{\prime}k_{2}.
\end{equation}
Supersaturation at the surface is defined as $log(x/x^{*})$. To illustrate the general behavior of this model, we have solved for
the time-dependence of the various quantities for the case that $a=a^{\prime}=0.04k_{0}$,
$k_{1}=0.02k_{0}$, $k_{1}^{\prime}=k_{2}^{\prime}=0.004k_{0}$, $k_{0}^{\prime}=0.5k_{0}$,
$k_{2}=0.01k_{0}$, $N=2$ with initial conditions $x_{0}(0)=1$, and
$x(0)=y(0)=c(0)=0$. Here the original time $t$ is normalized to $k_0t$. The value of $k_{0}^{\prime}$ reflects the idea that once adsorbed, a molecule spends on average a long time on the surface before being desorbed back to the bulk. The choice $k_{2} < k_{1}$ amounts to assuming that the activation barrier for aggregation alone is larger than that for adsorption and aggregation. Aside from these assumptions and the physical requirements that the
rates at which molecules join a cluster, $k_{1}$ and $k_{2}$, are
greater than the rates for the reverse processes and the detailed balance
condition, Eq.\ref{balance}, these values are somewhat flexible. Given an initially flat surface, $c_{N}(0)=0$,
islands will form causing the mass of the surface to increase until,
at long times, equilibrium is reached. At intermediate times, a more-or-less
steady growth regime is reached, see Fig. \ref{fig1}, corresponding
to a plateau in the growth velocity, $\frac{dc_{N}}{dt}$, which defines
a quasi-steady state growth velocity as shown in Fig. \ref{fig2}.
Taking the maximum growth velocity as a measure of the typical velocity
at intermediate times, one again finds a crossover between small $x$
and large $x$ regimes as illustrated in Figure \ref{fig3}.

\pagebreak{}

\section{One-dimensional Kinetic Monte Carlo model}

In order to study both the possibility of multiple growth regimes
without vanishing nucleation barrier and the roughening transition,
we have performed Kinetic Monte Carlo simulations of the growth of
a simple, one-dimensional surface in the spirit of the Solid-On-Solid
(SOS) model\cite{SOS}. The system consists of a set of
growth sites arranged as a one-dimensional lattice. There are $N$
sites, each characterized by a height $h_{i}$ where $i=0,...,N-1$.
Periodic boundaries are used so that formally, one has $h_{N}=h_{0}$.
The idea is that molecules randomly land on the surface at some specified
rate $\nu_{hit}$ representing the diffusion of molecules onto the
surface. Once on the surface, a molecule attempts to leave the surface
at a rate $\nu_{evap}$. Physically, $\nu_{hit}$ will be determined
by the concentration of the molecules in solution and the rate at
which they move while $\nu_{evap}$ is a measure of how fast a molecule
makes an attempt to leave the surface.

Nucleation is anomalous in one dimension due to the fact that the
{}``surface area'' of a cluster does not depend on its size. We
therefore introduce a rule designed to capture the most important
feature of surface nucleation which is the existence of a critical
cluster size. Specifically, molecules are not allowed to leave the
surface if they are part of a cluster that is of a specified critical
size or larger. The critical size $n_{c}$ is a parameter. Molecule
$j$ is ''part of a cluster'' of size $n$ if there are $n$ contiguous
sites, including site $j$, with heights the same as or greater than
$h_{j}$. Thus, small clusters can form spontaneously and dissociate
but a cluster at or above the critical size is stable.

A time step $dt=1/\left(\nu_{hit}+\nu_{evap}\right)$ is defined and
the dynamics of the kinetic Monte Carlo algorithm are as follows:
\begin{enumerate}
\item At time $t$ the system has some configuration of heights and total
mass $M\left(t\right)=\sum_{i=0}^{N}h_{i}$ and roughness $R\left(t\right)=\frac{1}{N}\sum_{i=0}^{N-1}\left|h_{i}-h_{i-1}\right|$.
Roughness could alternatively be characterized by the variance of
the heights, the correlation length, ...
\item The following is performed $N$ times:

\begin{enumerate}
\item Choose a random site, $j$.
\item choose a random number $u\in\left[0,1\right]$ and set $h_{j}\rightarrow h_{j}+1$
if $u<\nu_{hit}dt$.
\item if the height is not increased, (this happens with probability $1-$
$\nu_{hit}dt=\nu_{evap}dt$), then remove a molecule, $h_{j}\rightarrow h_{j}-1$ if site $j$ is not part of a supercritical cluster. 
\end{enumerate}
\item Set $t\rightarrow t+dt$.
\item Return to step $1$ until the desired number of cycles is completed. 
\end{enumerate}
As it stands, there are only two meaningful parameters: the ratio
of the two rates, $\nu_{hit}/\nu_{evap}$ and the size of the critical
cluster, $n_{c}$. Since super-critical clusters are absolutely stable 
there is no equilibrium state: clusters form no matter what value the parameters are given, as in the simplified form of the mean-field model discussed above.
The growth rate is calculated as $\left(M\left(t+dt\right)-M(t)\right)/dt$
and is normalized to the number of sites to give the corresponding growth velocity.
In the simulations, we also allow for the presence of {}``defects'',
localized regions in which the critical cluster size is smaller than
elsewhere, to serve as sites for heterogeneous nucleation.

Simulations have been performed under three circumstances: no defects,
a ''wall'' defect and a ''spiral'' defect. The {}``wall'' refers
to a set of sites, say sites $0,...,n_{c}$ for which the evaporation
rate is zero. These therefore grow very fast and serve as a source
for heterogeneous nucleation. For the ``spiral'' defect, the critical
nucleus for site $0$ is set to one, for site $1$ it is set to $1$,
and so forth up to site $n_{c}$. This is not meant to
  realistically model a spiral defect, but rather to test whether the
  results are sensitive to the shape and nature of the defect. The
  name indicates the similarity in shape of the resulting defect to a
  projection of a spiral defect onto two dimensions. In fact, we
  observe no qualitative difference in the results using the two
  different defects. The main effect is that at low supersaturation,
  $\nu_{hit}/\nu_{evap}<<1$, the rate of growth in the system with no
  defects is dominated by the time taken  for nucleation to occur  - by comparison, step growth happens relatively quickly. Thus,
the measured rate of increase of the solid is basically a measure
of the nucleation rate and is not directly relevant to the rate of
step growth. However, the same qualitative effects are seen with and
without defects. Intuitively, one expects that if $\nu_{hit}<<\nu_{evap}$ then nucleation
will be rare and growth will be dominated by attachment to existing
super-critical clusters, i.e. smooth step growth. When the hit rate
increases, nucleation becomes more probable until at some point nucleation
can occur easily and growth will be a combination of growth of existing
clusters and nucleation of new clusters. Simulations were performed
for a system consisting of 1500 sites and the velocity, defined as
the rate of change of the total mass, was averaged over 1000 cycles.
\begin{figure}
\begin{centering}
\resizebox{10cm}{!}{ {\includegraphics[width=14cm,angle=-90]{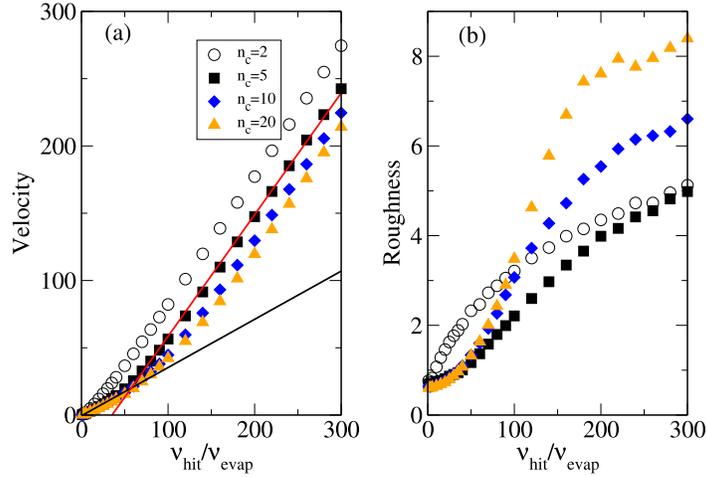}}} 
\par\end{centering}
\caption{Growth velocity $\frac{dM}{d\nu_{evap}t}$ (panel a) and roughness (panel b)  as functions of the hit rate for various choices of the
critical size, $n_{c}$ in the presence of a spiral defect.}
\label{fig4} 
\end{figure}

Figure \ref{fig4} shows the rate of growth of the surface as a function
of the hit rate for various choices of the size of the critical cluster
when a spiral defect is present. For a small critical cluster, $n_{c}=2$,
nucleation is almost instantaneous for all but the lowest values of $\nu_{hit}/\nu_{evap}$: growth is dominated by nucleation and the growth velocity increases more or less linearly with the
hit rate. For larger critical sizes, two different growth regimes
can be distinguished in the Figure. At small hit rates, growth is dominated by heterogeneous
nucleation at the defect with the subsequent growth of steps away
from the defect. This growth by aggregation is analogous to the $NY \rightarrow C_{N}$ scenario in the mean-field model. However, at high hit rates, nucleation becomes important
and the growth rate increases due to the combination of processes with nucleation dominating at high hit rates. This is analogous to the $(N-1)Y+X \rightarrow C_{N}$ process in the mean-field model.
The point of crossover between the regimes increases with increasing
critical size and  corresponds to a roughening transition
as is illustrated in Fig. \ref{fig4}b. Note that for the smallest critical cluster size, the transition region is too small to see in this figure. For critical clusters larger than $2$ and for small supersaturations,
the roughness is virtually independent of the critical cluster size
and increases slowly with the supersaturation while at higher supersaturation,
the transition to the rapid growth regime is also signaled by a dramatic
increase in roughness.

\begin{figure}

\begin{centering}
\resizebox{10cm}{!}{ {\includegraphics[height=10cm,angle=-90]{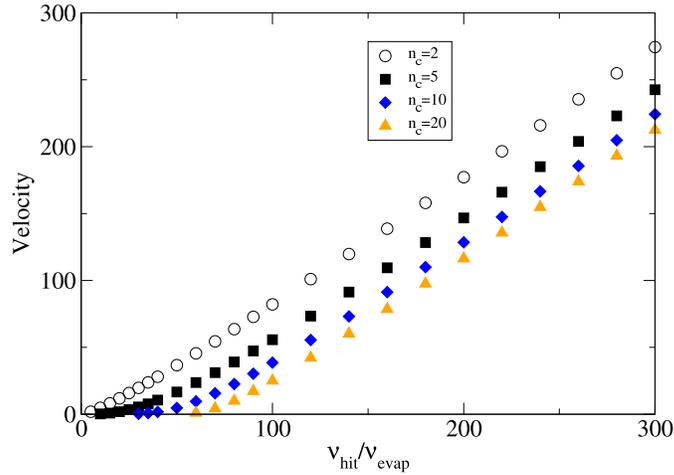}}} 
\par\end{centering}

\caption{Growth velocity as a function of hit rate for various choices of the
critical size, $n_{c}$ with no defect. The initial growth rate is
very small and is dominated by the nucleation rate.}

\label{fig6} 
\end{figure}

Finally, Fig. \ref{fig6} shows the growth rate when no defect is
present. The main difference from the previous results is that at
low supersaturations, the growth rate is very small since growth only
occurs after the nucleation of a super-critical cluster. Once such
a cluster forms, the growth of a new layer of crystal is quite rapid
so that the overall growth rate is dominated by the nucleation rate.
Nevertheless, a clear crossover from this slow growth regime, where
most mass is added by step growth, to a fast regime, where nucleation
plays an important role in adding mass, is evident. %

\section{Conclusions}

Within the context of classical nucleation theory, kinetic roughening is understood as
occurring when the supersaturation is so high that the size of the critical cluster becomes smaller than
one molecule\cite{Saito}. At low supersaturations, growth is dominated by the heterogeneous nucleation of 
steps that then grow smoothly to cover the crystal surface while at high supersaturation, above the transition, homogeneous nucleation
 also occurs leading to rapid growth of a rough surface. In this paper, the possibility of a similar transition \emph{for finite-sized critical clusters} has been investigated. Two models for the growth of crystal surfaces which show a transition
from a slow to a fast growth regime were presented. In the mean
field model, adsorption onto the surface with subsequent surface diffusion
and aggregation competes with direct formation of clusters to give
two different growth regimes for the formation of islands of new crystal.
At low bulk concentrations, the adsorption and aggregation mechanism
are dominant whereas at high bulk concentrations, the formation of
critical clusters by the formation of subcritical aggregates with
direct incorporation of material from solution dominates. The difference
in growth rates is greatest when the critical cluster is small (being
a factor of 2 for a critical size of 2). These results elaborate on
previous theoretical work that indicated the importance of direct
incorporation of material from the bulk\cite{Nicolis}.

A one-dimensional Kinetic Monte Carlo model based on the well-known
SOS model showed both different growth regimes and roughening. The
main difference from the usual SOS model is the definition of a critical
cluster size to simulate nucleation as it actually occurs on a two-dimensional
surface. This model showed a clear cross-over between a smooth, step-growth
dominated regime at low supersaturation and a rough, nucleation-dominated
regime at high supersaturation thus exhibiting the observed features
of the kinetic roughening transition. The growth velocities both below and above the transition are roughly linear functions of the hit rate (Fig. \ref{fig4}), or in physical terms, the concentration, just as in the mean-field model (see, e.g., Eqs.(\ref{result1}, \ref{result2})).

The main conclusion from this work is that the kinetic roughening-like
transitions do not necessarily require the vanishing of the critical
cluster size. Instead, {}``kinetic'' effects suffice to give qualitatively
different growth regimes and roughening of the surface. In future
work, it is intended to further elaborate the simple mean field model
presented here to give a semi-quantitative description of the results of simulation as well as of experiment. Finally, it is noteworthy that the experiments revealing the kinetic roughening mechanisms that motivated the present study\cite{Mike1} were
done using gel to quench convection which brings one close to microgravity conditions. It might be expected that a comparison under real microgravity conditions  would reveal further important features.

\begin{acknowledgments}
This work was supported by the European Space Agency under contract
number ESA AO-2004-070. 
\end{acknowledgments}
\bibliographystyle{apsrev} \bibliographystyle{apsrev}
%\bibliography{roughening}

\end{document}